\documentclass[aps,prd,twocolumn,letter]{revtex4}
\usepackage{color}
\usepackage{soul,color}
\usepackage{graphicx}
\usepackage{epsfig}

\usepackage{bm}
\newcommand{\be}{\begin{equation}}
\newcommand{\ee}{\end{equation}}
\newcommand{\ba}{\begin{eqnarray}}
\newcommand{\ea}{\end{eqnarray}}

\begin{document}
\title{Heavy - light flavour correlations of anisotropic flows at LHC energies within event-by-event transport approach} 
\author{Salvatore Plumari$^{a,b}$, Gabriele Coci$^{a,b}$, Vincenzo Minissale$^{b}$, Santosh K. Das$^{c}$, Yifeng Sun$^{b}$, Vincenzo Greco$^{a,b}$}

\affiliation{$^a$ Department of Physics and Astronomy 'Ettore Majorana', University of Catania, Via S. Sofia 64, 1-95125 Catania, Italy}
\affiliation{$^b$ Laboratori Nazionali del Sud, INFN-LNS, Via S. Sofia 62, I-95123 Catania, Italy}
\affiliation{$^{c}$ School of Physical Sciences, Indian Institute of Technology Goa, Ponda-403401, Goa,  India}

\date{}
\begin{abstract}
The heavy quarks (HQs) are unique probe of the hot QCD matter properties and their dynamics
is coupled to the locally thermalized expanding quark gluon plasma. We present here a novel study of the
event by event correlations between light and heavy flavour flow harmonics at LHC 
energy within a transport approach.
Interaction between heavy quarks and light quarks have been taken into account exploring the impact of different temperature dependence of the transport coefficients $D_s$ and $\Gamma$.
Our study indicates that $v^{heavy}_n-v^{light}_n$ correlation and the relative fluctuations
of anisotropic flows, $\sigma_{v_{n}}/\langle v_n \rangle$, are novel observables to understand the
heavy quark-bulk interaction and are sensitive to the temperature dependence 
even to moderate differences of $D_s(T)$, or $\Gamma(T)$. Hence a comparison of such new observables for HQ to upcoming experimental data at both RHIC and LHC
can put further constraints on heavy quark transport coefficients and in particular on its
temperature dependence toward a solid comparison between the
phenomenological determination and the lattice QCD calculations.

\vspace{2mm}

\end{abstract}
\maketitle
The main goal of the ongoing nucleus-nucleus collision programmes at Relativistic Heavy Ion Collider (RHIC) and Large
Hadron Collider (LHC) is to create and characterize a state of matter that behaves like a nearly perfect fluid having a 
remarkably small value of shear viscosity to entropy density ratio, $\eta/s=0.1$. The bulk properties of such a state of 
matter, called Quark Gluon Plasma (QGP)~\cite{Shuryak:2004cy,Jacak:2012dx}, are governed 
by the light quarks and gluons. 
Remarkable progress has been made towards the understanding of the properties of strongly
interacting QGP. Heavy quarks~\cite{Svetitsky:1987gq,Moore:2004tg,Dong:2019unq,Prino:2016cni,Andronic:2015wma,Aarts:2016hap,Cao:2018ews,Rapp:2018qla}, 
namely charm and bottom, thanks to their large masses, are considered as a solid probe
to characterize the QGP phase. They are produced in the early stage of the collisions and witness the entire 
space-time evolution of QGP and can act as an effective probe of the created matter.
Recently, phenomenological studies on heavy-hadron spectra and elliptic flows with different theoretical models
have been used to extract HQ diffusion coefficient at zero momentum with a determination of transport coefficient
which is within the current lattice QCD (lQCD) uncertainties \cite{Dong:2019unq,Prino:2016cni,Rapp:2018qla,Cao:2018ews}.
A next step would be to include more exclusive observables for the
determination of transport coefficients.
\newline
Heavy mesons nuclear suppression~\cite{Abelev:2006db,Adler:2005xv,Adam:2015sza}, $R_{AA}(p_T)$, the depletion 
of high $p_T$ hadrons produced in nucleus-nucleus collisions with respect to those produced in
proton-proton  collisions,  has been extensively used as probe of quark gluon plasma along
with the elliptic flow~\cite{Adare:2006nq,Abelev:2014ipa}, $v_2(p_T)=<cos(2\phi)>$, a measure of the anisotropy in the 
angular distribution of heavy mesons as a response to the initial anisotropy in coordinate space in non-central 
collisions. Several attempts have been made in this direction to study both these observables
theoretically to understand heavy quark dynamics in QGP
~\cite{vanHees:2005wb,vanHees:2007me,Gossiaux:2008jv,Das:2009vy,Alberico:2011zy,Uphoff:2012gb,Lang:2012cx,Song:2015sfa,Song:2015ykw,Das:2013kea,Cao:2015hia,Das:2015ana,Cao:2017hhk,Das:2017dsh,Sun:2019fud}.
In the recent past, a simultaneous study of both these observables, $R_{AA}(p_T)$ and $v_2(p_T)$,
received significant attention as it has the potential 
to constrain the temperature dependence of heavy quark transport 
coefficient in QGP ~\cite{Das:2015ana}. This is mainly due to the
different formation time of these observables that can probes different
stages of the fireball evolution.
The large elliptic flow~\cite{Adare:2006nq,Abelev:2014ipa}, $v_2$ , and small nuclear suppression 
factor~\cite{Abelev:2006db,Adler:2005xv,Adam:2015sza}, $R_{AA}$ , of the heavy flavour mesons observed
at RHIC and LHC colliding energy are considered as an indication that the heavy quark interact strongly with the bulk medium. 
It has been shown, in ref ~\cite{Plumari:2011mk}, that quasi particle model (QPM) is able to reproduce the lattice QCD 
pressure, energy density and interaction measure $T^\mu_\mu=\epsilon-3 P$. The main feature of this QPM approach 
is that the resulting coupling is significantly stronger than the one coming from pQCD running coupling, particularly 
at $T \rightarrow T_c$. This feature of QPM has been observed by other groups \cite{Berrehrah:2014kba} also
when quasi-particle widths (off-shell dynamics) are accounted for.
In fact, the recent success of QPM to describe the heavy quark-bulk interaction in the QGP has been credited to the large enhancement of the coupling near $T_c$. 
A similar mechanism acts also in the T-matrix approach of TAMU \cite{vanHees:2005wb,vanHees:2007me,He:2012df, He:2011qa}
and the PHSD transport of Frankfurt \cite{Song:2015ykw} and more recently has been investigated by the LBL-CCNU collaboration \cite{Cao:2016gvr}.
Therefore, temperature dependence of heavy quark transport coefficients is a key ingredient 
for the simultaneous description of heavy meson  $R_{AA}(p_T)$ and $v_2(p_T)$. This matter has also been observed 
in the high $p_T$ range and in the light sector ref. \cite{SCARDINA,Liao:2008dk, Xu:2014tda}, independently. Recent studies based 
on T-dependence K factor ~\cite{Cao:2016gvr} and 
Bayesian model-to-data analysis~\cite{Xu:2017obm} also obtained similar conclusions from different view points.
This along with an hadronization via coalescence plus 
fragmentation is the main underlying ingredient for a simultaneous description of heavy quark $R_{AA}$ and $v_{2}$ observed 
experimentally~\cite{Scardina:2017ipo}.
Most of the studies of $v_2$ of HF have been pursued discarding the more realistic dynamics on an event-by-event fluctuation framework with some exception.
Recently, the triangular flow $v_3$ has been investigated in theoretical studies based on Langevin approach on an event-by-event analysis
~\cite{Nahrgang:2014vza, Prado:2016szr, Nahrgang:2016wig,Beraudo:2018tpr,Katz:2019fkc}.
The present work is an extension of our recent works presented 
in Ref.~\cite{Das:2013kea, Das:2015ana, Scardina:2017ipo} introducing an event-by-event 
fluctuating initial condition~\cite{Plumari:2015cfa, Plumari:2019gwq} 
which allow us to study also the odd harmonics. 
However, the main aim is to focus for the
first time on the $v^{heavy}_n-v^{light}_n$ correlations up to 5th order.
We show that within this approach with
hadronization by coalescence plus fragmentation we are able to describe the experimental data not only for $v_1$ and $v_2$ of D mesons at LHC energy but also $v_3$ for different collision centralities as done in previous works 
~\cite{Das:2013kea,Das:2015ana,Das:2016cwd,Scardina:2017ipo,Plumari:2017ntm}.
We also study the correlations that takes place between the heavy quarks and the
bulk of quarks and gluons. Finally, we show how the study of heavy-light correlations and the $v_n$ distributions can
give information about the heavy quark interaction with the QGP. In particular we show that the linear correlation coefficient of 
$v_n^{heavy}-v_n^{light}$ (a measure of the degree of correlation) and the relative variance $\sigma_{v_{n}}/\langle v_n \rangle$ are observables highly sensitive to the
temperature dependence of the transport coefficients.
\newline
The momentum evolution of the charm quark distribution function in QGP is obtained by solving the relativistic Boltzmann 
transport equations~\cite{Ferini:2008he, Scardina:2017ipo}: 
\ba
& & p^{\mu} \partial_{\mu}f_{Q}(x,p)= {\cal C}[f_q,f_g,f_{Q}](x,p) \nonumber  \\
& & p^{\mu}_q \partial_{\mu}f_{q}(x,p)= {\cal C}[f_q,f_g](x_q,p_q)  \nonumber  \\
& & p^{\mu}_g \partial_{\mu}f_{g}(x,p)= {\cal C}[f_q,f_g](x_g,p_g)   
\label{B_E} 
\ea
where $f_i(x,p)$ is the on-shell phase space one-body distribution function for the $i$ parton and
${\cal{C}}[f_q, f_g, f_{Q}](x,p)$ is the relativistic Boltzmann-like collision integral. The phase-space
distribution function of the bulk medium consists of quark and gluons entering the equation for charm quarks as
an external quantities in ${\cal{C}}[f_q,f_g,f_{Q}]$.
We assume that the evolution of $f_q$ and $f_g$ are independent of $f_{Q}(x,p)$ and discard collisions between charm quarks
which is by far a solid approximation.  
We are interested in the evolution of the HQ distribution function $f_{Q}(x,p)$ scattering with the bulk medium. The evolution of the bulk of 
quark and gluons is 
instead given by the solution of the other two transport equations where the ${\cal C}[f_q,f_g]$ is 
tuned to a fixed $\eta/s(T)$, as discussed
in detail in ref.~\cite{Ruggieri:2013ova}.
\newline
For the modelling of the bulk we adopted the same approach described in Ref.s \cite{Plumari:2015cfa, Plumari:2019gwq}.
In order to set the initial geometry the nucleons within the two nuclei have been distributed according to a Woods-Saxon
distribution. In this way a discrete distribution for these nucleons is generated.
The geometrical method is used to determine if the two nucleons, one from the nucleus A and the other
one from the nucleus B, are colliding. The two nucleons collide if the relative distance in the transverse
plane is $d_T \le \sigma_{NN}/\pi$ where $\sigma_{NN}$ is the nucleon-nucleon cross section.
A $\sigma_{NN}=7.0 fm^2$ was employed in our calculations. The discrete distribution for the nucleons
is converted into a smooth one by assuming for each nucleon a gaussian distribution centered in
the nucleon position.
Finally we convert the nucleon distribution into the parton density distribution in transverse plane
$\rho_T(x,y)$ which is given by $\rho_T(x,y) = K \sum_{i=1}^{N_{part}} \exp{\bigg[-((x-x_i)^2+(y-y_i)^2)/(2\sigma_{xy}^2)\bigg]}$
the proportionality coefficient $K$ is fixed by the experimental longitudinal distribution $dN/dy$ while 
$\sigma_{xy}$ is the Gaussian width which regulates the smearing of the fluctuations and has been fixed to $\sigma_{xy} = 0.5 \, fm$ as done also in the hydrodynamics
framework \cite{Qin:2010pf,Petersen:2010cw}. In our calculation we have assumed initially a longitudinal boost invariant distribution from $y=-2.5$ to $y=2.5$.
We initialize the charm quark distribution  in the coordinate space in accordance with the number 
of binary nucleon-nucleon collisions, $N_{coll}$, from the Monte Carlo Glauber model. 
For the momentum distribution we use charm quark production in 
Fixed Order + Next - to - Leading Log (FONLL)~\cite{Cacciari:2012ny} 
which describes the D-meson spectra in proton-proton collisions after fragmentation. For detail we refer to 
our earlier work in Ref.~\cite{Scardina:2017ipo}. 
For the heavy quark bulk interaction we consider a QPM whose main feature is that the resulting 
coupling is significantly stronger than the one coming from pQCD running coupling, particularly as $T\rightarrow T_c$.
The scattering matrix ${\cal M}_{{(q,g)+Q} \leftrightarrow {(q,g)+Q}}$  have been evaluated
considering the leading-order diagram with the effective coupling $g(T)$ that leads to effective vertices and a
dressed massive gluon propagator for $gQ\leftrightarrow gQ$ and massive quark propagator for $qQ \leftrightarrow qQ$ scatterings. 
For the scattering matrix we are using the Combridge matrix~\cite{Combridge:1978kx} that includes $s$, $t$, $u$ channel and their interference terms. 
For detail we refer to our earlier work~\cite{Das:2015ana, Scardina:2017ipo}. 
Once the temperature of the QGP phase goes below the quark-hadron transition temperature, $T_c=155$ MeV, we hadronize 
the charm quark to D-meson. For heavy quark hadronization, we consider an hybrid approach of hadronization by coalescence
plus fragmentation. We use Peterson fragmentation function~\cite{Peterson:1982ak}:
$f(z) \propto \lbrack z \lbrack 1- \frac{1}{z}- \frac{\epsilon_c}{1-z} \rbrack^2 \rbrack^{-1}$
%
where $z=p_D/p_c$ is the momentum fraction of the D-meson fragmented from the charm quark and
$\epsilon_c$ is a free parameter to fix the shape of the fragmentation function, so that one 
can describe D-meson production in proton-proton collisions \cite{Moore:2004tg}. For the coalescence we use a model where the particle production is given by the coalescence integral based on a Wigner formalism, where the widths of the Wigner function
are fixed by the mean square radius of the D meson, for detail we refer to our earlier work~\cite{Plumari:2017ntm}.
\begin{figure}[t]
\begin{center}
\includegraphics[width=14pc,clip=true]{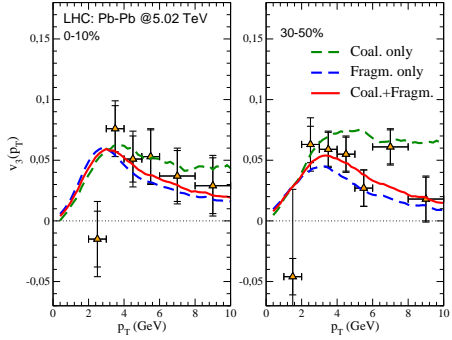}
\caption{$D^{0}$ meson $v_3(p_T)$ at mid-rapidity for the two centralities classes ($0-10\%$) (left)
and ($30-50\%$) (right).
Green dashed line correspond to D mesons produced only via coalescence while blue dashed line via only fragmentation. The red solid line refer to the case with both coalescence plus fragmentation. Data taken from \cite{Sirunyan:2017plt}.}
\label{FIG:vnpT}
\end{center}
\end{figure}
In Fig.\ref{FIG:vnpT} are shown the results for the triangular flows as a function of momentum in $Pb+Pb$ collisions 
at $\sqrt{s}=5.02 \,\rm ATeV$. We show explicitly the different contributions of different hadronization processes allowing a direct access to the role played by coalescence and fragmentation.
The blue dashed lines indicate the $v_3(p_T)$ for D mesons that we obtain considering only the fragmentation
as hadronization mechanism while the green dashed line correspond to the case via only coalescence.
The $v_n$ developed by only coalescence is larger than the $v_n$ developed by only fragmentation.
This difference originates from the fact that the D meson anisotropic flows formed via coalescence reflects both the heavy 
quark and the bulk anisotropies in momentum space and this leads to an enhancement of the $v_n(p_T)$. 
Finally, the solid red line shows the $v_n(p_T)$ of  D mesons produced via coalescence plus fragmentation.
The underlying fraction of coalescence and fragmentation is the one that allows a good description of D meson spectra and the recently measured $\Lambda_c/D^0$ ratio at RHIC and LHC \cite{Adam:2019hpq,Vermunt:2019ecg}, see Ref. \cite{Plumari:2017ntm}. 
\begin{figure}[t]
\begin{center}
  \includegraphics[width=14pc,clip=true]{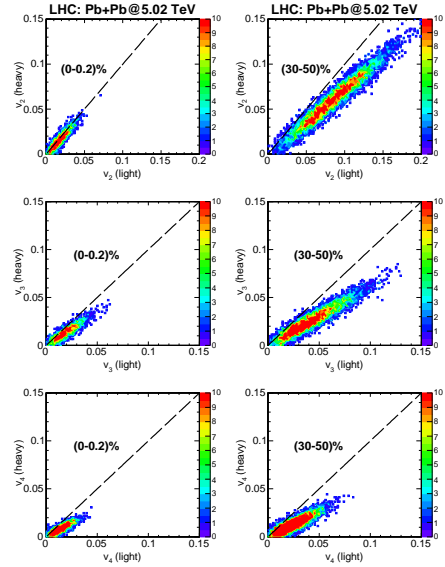}
\caption{Event-by-event correlation between $v_n^{(heavy)}$ and $v_n^{(light)}$ for $n=2,3,4$ for $Pb+Pb$ collisions at $\sqrt{s}=5.02 TeV$ for ($30-50 \%$) and ($0-0.2\%$) centrality cuts.}
\label{FIG:vnvn}
\end{center}
\end{figure}
In recent years, the correlation between integrated $v_2$ and high order harmonics $v_3, \, v_4$ for light hadrons with the 
initial asymmetry in coordinate space $\epsilon_2, \epsilon_3$ and $\epsilon_4$ have been studied in the event-by-event ideal and viscous 
hydrodynamics and transport framework \cite{Gardim:2011xv,Chaudhuri:2012mr,Niemi:2012aj,Plumari:2015cfa}.
In this paper we extend these studies to the heavy quark sector with the main novelty of considering the correlations between charm quarks and light quarks within an event-by-event transport approach.
In Fig.~\ref{FIG:vnvn}, it has been shown the two-dimensional plots of the integrated flow $v_n^{light}$ for light quarks 
as a function of the corresponding final integrated flow coefficients $v_n^{heavy}$ for heavy quarks .
The results are for $Pb+Pb$ collisions at $\sqrt{s_{NN}}=5.02 \, TeV$ for two different centralities $(0-0.2)\%$ and $(30-50)\%$.
The viscosity has been fixed to $4\pi\eta/s=1$ plus a kinetic f.o. realized by the increase in $\eta/s(T)$ for more details see
\cite{Plumari:2015cfa,Plumari:2019gwq}.
As shown in the upper panel we observe a strong linear correlation between $v_2^{(heavy)}$ and $v_2^{(light)}$ at both central and peripheral collisions.
For the higher harmonics, we observe a reduction of the 
linear correlation in comparison to the second harmonic.
\begin{figure}[t]
\begin{center}
  \includegraphics[width=14pc,clip=true]{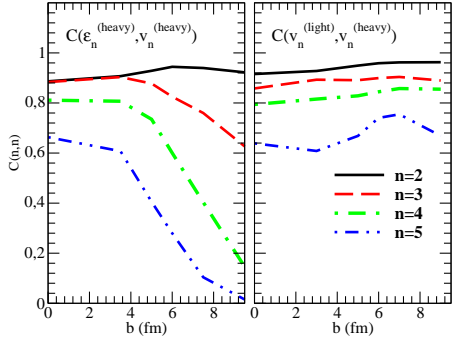}
  \caption{Left panel: correlation coefficient between the initial $\epsilon_n$ and $v_n$ of charm quarks at different impact parameters.
    Right panel: correlation coefficient between heavy quarks $v_n$ and light quarks $v_n$ at different impact parameters.
    Different colours are for different harmonics. These results have been obtained within QPM.}
\label{Cnn}
\end{center}
\end{figure}
A measure of the linear correlation is given by the correlation coefficient 
$C(n,m)$ expressed as:
\begin{equation}
C(n,m)=\frac{\sum_{i}(v_n^{L,i}-\langle v_n^{L} \rangle)(v_m^{H,i}-\langle v_m^{H}\rangle)}{\sqrt{\sum_{i}(v_n^{L,i}-\langle v_n^{L} \rangle)^2\sum_{i}(v_m^{H,i}-\langle v_m^{H} \rangle)^2}}
\end{equation}
where $v_n^{L,i}$ and $v_m^{H,i}$ are the values of anisotropic flows corresponding to the event $i$ and
respectively for light and heavy quarks.
A $C(n,m) \approx 1$ corresponds to a strong linear correlation between light and heavy anisotropic flows.
The results shown in this section have been obtained with a number of event $N_{event}=5000$ for each centrality class and a 
total number of test particle per event $N_{test}=2 \cdot 10^{6}$.
\newline
In the left panel of Fig.~\ref{Cnn} we show the $C(n,n)$ between the initial $\epsilon_n$ and final $v_n$ of charm quarks as a function of impact parameter in the left panel while in the right panel the correlation coefficient between final heavy quarks $v_n^{(heavy)}$ and light quarks $v_n^{(light)}$.
As shown in the left panel the linear correlation coefficient is a decreasing function with respect the impact parameter for all the harmonics.
Moreover, we observe that only $v_2$ is strongly correlated with the corresponding initial eccentricities $\epsilon_2$. 
On the other hand, in the right panel of Fig.~\ref{Cnn} (black solid line), the correlation coefficient $C(2,2)\approx 0.95$ and it remains almost flat 
and independent of impact parameter. 
We observe for $v_2$ and $v_3$ approximatively the same degree of correlation
while a lower degree of correlation it is shown for higher harmonics 
$n=4,5$.
However, the main original finding is that the $v_n^{(heavy)}$ remains at least up to $n=4$ strongly correlated to $v_n^{(light)}$  at different impact parameter at variance with the correlation to the initial $\epsilon_n$. 
This means that the building up of $v_n^{(heavy)}$ is driven by the $v_n$ of the bulk 
while the correlation to the initial eccentricities is nearly lost for $b \ge 5 fm$. 
This suggests that for non central collisions a strong correlation between heavy quarks $v_n$ and light quarks $v_n$ originates from the heavy quarks and bulk interaction. 
Of course the experimental observation of these patterns would give a strong confirmation that the mechanism of $v_2$ build-up is the one essentially underlying most of the present theoretical description. 
\begin{figure}[t]
\begin{center}
  \includegraphics[width=12pc,clip=true]{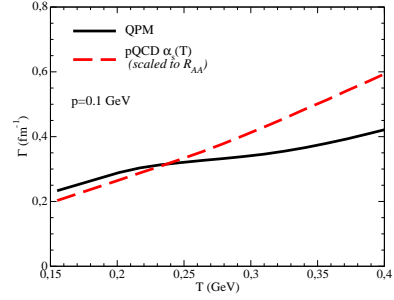}
  \caption{Temperature dependence of the drag coefficient $\Gamma$. Black solid line corresponds to QPM. The red dashed lines 
  correspond to pQCD interaction scaled in order to reproduce the sane $R_{AA}$ of QPM for the case with $\alpha_s(T)$.}
\label{FIG:Drag}
\end{center}
\end{figure}
In addition to this we will address the effect of the temperature dependence of the interaction, namely the drag coefficient on  both correlation and $v_n$ distribution.
There have been several studies of the T dependence, and it has been shown that it is an important aspect to be able to reproduce both $R_{AA}$ and $v_2$. 
However at present other aspects like the initial conditions, the details of the bulk expansion, or the details of the hadronization can reflect into a significant 
uncertainty of the T dependence of $\Gamma(T)$ or $D_s(T)$.
In Fig.~\ref{FIG:Drag} we show the behaviour of the drag coefficient $\Gamma$ with temperature in QPM by black solid line.
By red dashed line we show the same coefficient obtained
within the framework of pQCD with a temperature dependent ${\alpha}_{s}(T)$
and where we use the same bulk
used for QPM but we upscale the interaction of pQCD case in order to reproduce the same $R_{AA}(p_T)$ obtained in the QPM case.
Here we present a study of the impact of a moderate difference in the T dependence of $\Gamma(T)$ exploring a range that is even narrower than the current uncertainties \cite{Dong:2019unq,Rapp:2018qla}. The $\Gamma(T)$ has been tuned to reproduce the same $R_{AA}(p_T)$ of D mesons and exhibit a similar $v_2(p_T)$. 
\begin{figure}[t]
\begin{center}
\includegraphics[width=12pc,clip=true]{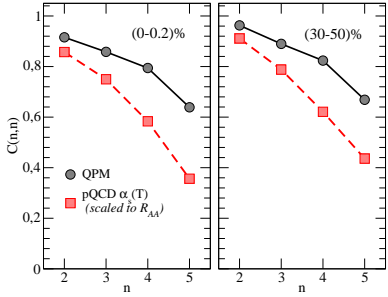}
\caption{Event-by-event correlation coefficient between D-meson $v_n$ and light hadron $v_n$ as a function of 
the order of the harmonic $n$ obtained within QPM (solid line) and pQCD (dashed line).} 
\label{Cnn1}
\end{center}
\end{figure}
In Fig.~\ref{Cnn1} we present the correlation coefficient $C(n,n)$ as a function of the order of the harmonic $n$. 
We have computed $C(n,n)$ for both QPM and pQCD to understand the impact of 
temperature dependence of heavy quark transport coefficients on the heavy-light $v^{HF}_2-v^{LF}_2$ correlation 
coefficients. As shown, the correlation coefficient decrease versus to the order of harmonics for both 
pQCD and QPM as expected from Fig.~\ref{Cnn}. However, the  correlation is stronger for 
QPM than pQCD for all the harmonics considered in this manuscript for both centralities. This highlights 
the impact of temperature dependence of heavy quark transport coefficients on event-by-event D-meson $v_n$ and 
light hadron $v_n$ correlation coefficients especially for semi-peripheral collisions.
Notice that different harmonics have different formation time \cite{Plumari:2015cfa}. 
This implies that for different harmonics their correlation coefficients result more sensitive to the temperature dependence of the transport coefficients.
In fact we see that the QPM $\Gamma(T)$ induce a stronger $C(n,n)$ because it has a larger drag at $T\le 1.5 T_c$ where most of $v_n$ develop and the effect increase with $n$.
Therefore this study suggests that the comparison of the theoretical data with the upcoming experimental results will give a more
powerful constraint on the T dependence of the HQ transport coefficient considering that moderate difference in $\Gamma(T)$ like those shown in Fig.\ref{FIG:Drag} can induce quite different $C(n,n)$. \newline
We finalize the present study by discussing also the normalized $v_n$ variances.
%
\begin{figure}[t]
\begin{center}
\includegraphics[width=14pc,clip=true]{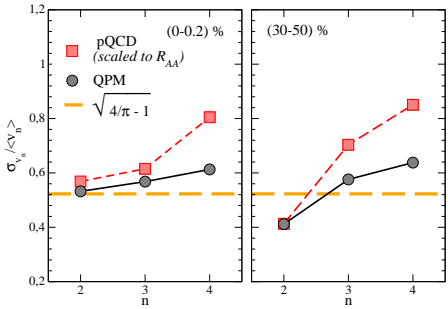}\hspace{2pc}
\caption{$\sigma_{v_{n}}/\langle v_n \rangle$ as a function of the order of harmonics $n$ for QPM (solid lines) and pQCD (dashed line). 
The orange dashed lines indicate the value $\sqrt{4/\pi -1}$ expected for a 2D Gaussian distribution.
}
\label{Fig:sigmavn}
\end{center}
\end{figure}
In fact some interesting properties of heavy quarks $v_n$ distributions can be inferred by studying the 
centrality dependence of the relative fluctuations $\sigma_{v_{n}}/\langle v_n \rangle$
where $\sigma_{v_{n}}$ are the standard deviation for $v_n$.
In Fig.\ref{Fig:sigmavn} it is shown the ratios $\sigma_{v_{n}}/\langle v_n \rangle$ for two different
centralities and for the two cases discussed above. 
As shown in Fig.\ref{Cnn} the second and third harmonics are the most correlated ones
$C(2,2) \approx C(3,3) \approx 0.9$ therefore also the $\langle v_n \rangle$ distributions $p(v_n)$
tend to reflect the $p(v_n)$ distributions of the bulk which are strongly correlated to the initial
eccentricity distributions.
The $\sigma_{v_{2}}/\langle v_2 \rangle$ is a decreasing function with the centrality
while $\sigma_{v_{3}}/\langle v_3 \rangle$ is almost independent to centrality for the QPM T dependence and
they are very close to the the value $\sqrt{4/\pi-1}$ shown by the dashed lines.
These results imply that the distributions of $v_3$ are consistent with the
fluctuation-only scenario discussed in \cite{Aad:2013xma} while the distribution of
$v_2$ is close to this limit for most central collisions while for mid-peripheral
collisions the impact of global average geometry decreases the fluctuations. 
If one looks at the evolution of $\sigma_{v_{n}}/\langle v_n \rangle$
with $n$ it emerges a strong sensitivity to the T dependence of $\Gamma$. This constitute
a further novel way to get information on the charm interaction. We also notice that for charm quark seems possible
to have normalized variance for $n \ge 3$ that are even quite large than the gaussian fluctuations only scenario
(shown by dashed lines in Fig.\ref{Fig:sigmavn}). 
Therefore also $\sigma_{v_{n}}/\langle v_n \rangle$ for $n > 2$ are 
quite strongly sensitive observables of the temperature dependence of the transport coefficients. 
\newline
In summary,  we have studied heavy-light $v^{HF}_2-v^{LF}_2$ correlation in event-by-event up to 5th order. 
Heavy quark bulk interaction has been treated with both QPM and pQCD.  For heavy quark momentum evolution 
we have solved the relativistic Boltzmann transport equation. For the bulk initialization, we introduce an event-by-event fluctuating
initial condition which allow us to access the study of the odd harmonics.
We have evaluated the heavy-light correlation coefficients $C(v_n^{(heavy)},v_n^{(light)})$ as a function of impact parameter and 
order of the harmonic $n$. 
We see that for $n \le 4$ it increases with the order of the harmonics and as a function of centrality show a nearly flat behaviour that, if observed experimentally, 
will be a signature that the HQ anisotropies are driven by the bulk ones.
Moreover, we have mainly found a strong impact of temperature dependence 
of heavy quark transport coefficients on $C(v_n^{(heavy)},v_n^{(light)})$ and $\sigma_{v_{n}}/\langle v_n \rangle$ as a function of the order of harmonics $n$. 
Comparison of results presented in this manuscript with the upcoming experimental 
data will help to constrain heavy quark transport coefficients and disentangle difference energy loss model.
\vspace{2mm}
\section*{Acknowledgments}
S.P. , V.M. and V.G. acknowledge the support by INFN-SIM national project and linea di intervento 2, DFA-Unict.
S.K.D. acknowledges the support by the National Science Foundation of China (Grants No.11805087 and No. 11875153).
%
%
%

\begin{thebibliography}{63}
\expandafter\ifx\csname natexlab\endcsname\relax\def\natexlab#1{#1}\fi
\expandafter\ifx\csname bibnamefont\endcsname\relax
  \def\bibnamefont#1{#1}\fi
\expandafter\ifx\csname bibfnamefont\endcsname\relax
  \def\bibfnamefont#1{#1}\fi
\expandafter\ifx\csname citenamefont\endcsname\relax
  \def\citenamefont#1{#1}\fi
\expandafter\ifx\csname url\endcsname\relax
  \def\url#1{\texttt{#1}}\fi
\expandafter\ifx\csname urlprefix\endcsname\relax\def\urlprefix{URL }\fi
\providecommand{\bibinfo}[2]{#2}
\providecommand{\eprint}[2][]{\url{#2}}

\bibitem[{\citenamefont{Shuryak}(2005)}]{Shuryak:2004cy}
\bibinfo{author}{\bibfnamefont{E.~V.} \bibnamefont{Shuryak}},
  \bibinfo{journal}{Nucl. Phys.} \textbf{\bibinfo{volume}{A750}},
  \bibinfo{pages}{64} (\bibinfo{year}{2005})

\bibitem[{\citenamefont{Jacak and Muller}(2012)}]{Jacak:2012dx}
\bibinfo{author}{\bibfnamefont{B.~V.} \bibnamefont{Jacak}} \bibnamefont{and}
  \bibinfo{author}{\bibfnamefont{B.}~\bibnamefont{Muller}},
  \bibinfo{journal}{Science} \textbf{\bibinfo{volume}{337}},
  \bibinfo{pages}{310} (\bibinfo{year}{2012}).

\bibitem[{\citenamefont{Svetitsky}(1988)}]{Svetitsky:1987gq}
\bibinfo{author}{\bibfnamefont{B.}~\bibnamefont{Svetitsky}},
  \bibinfo{journal}{Phys. Rev.} \textbf{\bibinfo{volume}{D37}},
  \bibinfo{pages}{2484} (\bibinfo{year}{1988}).

\bibitem[{\citenamefont{Moore and Teaney}(2005)}]{Moore:2004tg}
\bibinfo{author}{\bibfnamefont{G.~D.} \bibnamefont{Moore}} \bibnamefont{and}
  \bibinfo{author}{\bibfnamefont{D.}~\bibnamefont{Teaney}},
  \bibinfo{journal}{Phys. Rev.} \textbf{\bibinfo{volume}{C71}},
  \bibinfo{pages}{064904} (\bibinfo{year}{2005})

\bibitem[{\citenamefont{Dong and Greco}(2019)}]{Dong:2019unq}
\bibinfo{author}{\bibfnamefont{X.}~\bibnamefont{Dong}} \bibnamefont{and}
  \bibinfo{author}{\bibfnamefont{V.}~\bibnamefont{Greco}},
  \bibinfo{journal}{Prog. Part. Nucl. Phys.} \textbf{\bibinfo{volume}{104}},
  \bibinfo{pages}{97} (\bibinfo{year}{2019}).

\bibitem[{\citenamefont{Prino and Rapp}(2016)}]{Prino:2016cni}
\bibinfo{author}{\bibfnamefont{F.}~\bibnamefont{Prino}} \bibnamefont{and}
  \bibinfo{author}{\bibfnamefont{R.}~\bibnamefont{Rapp}}, \bibinfo{journal}{J.
  Phys.} \textbf{\bibinfo{volume}{G43}}, \bibinfo{pages}{093002}
  (\bibinfo{year}{2016})

\bibitem[{\citenamefont{Andronic et~al.}(2016)}]{Andronic:2015wma}
\bibinfo{author}{\bibfnamefont{A.}~\bibnamefont{Andronic}}
  \bibnamefont{et~al.}, \bibinfo{journal}{Eur. Phys. J.}
  \textbf{\bibinfo{volume}{C76}}, \bibinfo{pages}{107} (\bibinfo{year}{2016})

\bibitem[{\citenamefont{Aarts et~al.}(2017)}]{Aarts:2016hap}
\bibinfo{author}{\bibfnamefont{G.}~\bibnamefont{Aarts}} \bibnamefont{et~al.},
  \bibinfo{journal}{Eur. Phys. J.} \textbf{\bibinfo{volume}{A53}},
  \bibinfo{pages}{93} (\bibinfo{year}{2017})

\bibitem[{\citenamefont{Cao et~al.}(2019)}]{Cao:2018ews}
\bibinfo{author}{\bibfnamefont{S.}~\bibnamefont{Cao}} \bibnamefont{et~al.},
  \bibinfo{journal}{Phys. Rev.} \textbf{\bibinfo{volume}{C99}},
  \bibinfo{pages}{054907} (\bibinfo{year}{2019})

\bibitem[{\citenamefont{Beraudo et~al.}(2018)}]{Rapp:2018qla}
\bibinfo{author}{\bibfnamefont{A.}~\bibnamefont{Beraudo}} \bibnamefont{et~al.},
  \bibinfo{journal}{Nucl. Phys.} \textbf{\bibinfo{volume}{A979}},
  \bibinfo{pages}{21} (\bibinfo{year}{2018})

\bibitem[{\citenamefont{Abelev et~al.}(2007)}]{Abelev:2006db}
\bibinfo{author}{\bibfnamefont{B.~I.} \bibnamefont{Abelev}}
  \bibnamefont{et~al.} (\bibinfo{collaboration}{STAR}), \bibinfo{journal}{Phys.
  Rev. Lett.} \textbf{\bibinfo{volume}{98}}, \bibinfo{pages}{192301}
  (\bibinfo{year}{2007}), \bibinfo{note}{[Erratum: Phys. Rev.
  Lett.106,159902(2011)]}

\bibitem[{\citenamefont{Adler et~al.}(2006)}]{Adler:2005xv}
\bibinfo{author}{\bibfnamefont{S.~S.} \bibnamefont{Adler}} \bibnamefont{et~al.}
  (\bibinfo{collaboration}{PHENIX}), \bibinfo{journal}{Phys. Rev. Lett.}
  \textbf{\bibinfo{volume}{96}}, \bibinfo{pages}{032301}
  (\bibinfo{year}{2006})

\bibitem[{\citenamefont{Adam et~al.}(2016)}]{Adam:2015sza}
\bibinfo{author}{\bibfnamefont{J.}~\bibnamefont{Adam}} \bibnamefont{et~al.}
  (\bibinfo{collaboration}{ALICE}), \bibinfo{journal}{JHEP}
\textbf{\bibinfo{volume}{03}}, \bibinfo{pages}{081} (\bibinfo{year}{2016})

\bibitem[{\citenamefont{Adare et~al.}(2007)}]{Adare:2006nq}
\bibinfo{author}{\bibfnamefont{A.}~\bibnamefont{Adare}} \bibnamefont{et~al.}
  (\bibinfo{collaboration}{PHENIX}), \bibinfo{journal}{Phys. Rev. Lett.}
  \textbf{\bibinfo{volume}{98}}, \bibinfo{pages}{172301}
  (\bibinfo{year}{2007})

\bibitem[{\citenamefont{Abelev et~al.}(2014)}]{Abelev:2014ipa}
\bibinfo{author}{\bibfnamefont{B.~B.} \bibnamefont{Abelev}}
  \bibnamefont{et~al.} (\bibinfo{collaboration}{ALICE}),
  \bibinfo{journal}{Phys. Rev.} \textbf{\bibinfo{volume}{C90}},
  \bibinfo{pages}{034904} (\bibinfo{year}{2014})

\bibitem[{\citenamefont{van Hees et~al.}(2006)\citenamefont{van Hees, Greco,
  and Rapp}}]{vanHees:2005wb}
\bibinfo{author}{\bibfnamefont{H.}~\bibnamefont{van Hees}},
  \bibinfo{author}{\bibfnamefont{V.}~\bibnamefont{Greco}}, \bibnamefont{and}
  \bibinfo{author}{\bibfnamefont{R.}~\bibnamefont{Rapp}},
  \bibinfo{journal}{Phys. Rev.} \textbf{\bibinfo{volume}{C73}},
  \bibinfo{pages}{034913} (\bibinfo{year}{2006})

\bibitem[{\citenamefont{van Hees et~al.}(2008)\citenamefont{van Hees,
  Mannarelli, Greco, and Rapp}}]{vanHees:2007me}
\bibinfo{author}{\bibfnamefont{H.}~\bibnamefont{van Hees}},
  \bibinfo{author}{\bibfnamefont{M.}~\bibnamefont{Mannarelli}},
  \bibinfo{author}{\bibfnamefont{V.}~\bibnamefont{Greco}}, \bibnamefont{and}
  \bibinfo{author}{\bibfnamefont{R.}~\bibnamefont{Rapp}},
  \bibinfo{journal}{Phys. Rev. Lett.} \textbf{\bibinfo{volume}{100}},
  \bibinfo{pages}{192301} (\bibinfo{year}{2008})

\bibitem[{\citenamefont{Gossiaux and Aichelin}(2008)}]{Gossiaux:2008jv}
\bibinfo{author}{\bibfnamefont{P.~B.} \bibnamefont{Gossiaux}} \bibnamefont{and}
  \bibinfo{author}{\bibfnamefont{J.}~\bibnamefont{Aichelin}},
  \bibinfo{journal}{Phys. Rev.} \textbf{\bibinfo{volume}{C78}},
  \bibinfo{pages}{014904} (\bibinfo{year}{2008})

\bibitem[{\citenamefont{Das et~al.}(2009)\citenamefont{Das, Alam, and
  Mohanty}}]{Das:2009vy}
\bibinfo{author}{\bibfnamefont{S.~K.} \bibnamefont{Das}},
  \bibinfo{author}{\bibfnamefont{J.-e.} \bibnamefont{Alam}}, \bibnamefont{and}
  \bibinfo{author}{\bibfnamefont{P.}~\bibnamefont{Mohanty}},
  \bibinfo{journal}{Phys. Rev.} \textbf{\bibinfo{volume}{C80}},
  \bibinfo{pages}{054916} (\bibinfo{year}{2009})

\bibitem[{\citenamefont{Alberico et~al.}(2011)\citenamefont{Alberico, Beraudo,
  De~Pace, Molinari, Monteno, Nardi, and Prino}}]{Alberico:2011zy}
\bibinfo{author}{\bibfnamefont{W.~M.} \bibnamefont{Alberico}},
  \bibinfo{author}{\bibfnamefont{A.}~\bibnamefont{Beraudo}},
  \bibinfo{author}{\bibfnamefont{A.}~\bibnamefont{De~Pace}},
  \bibinfo{author}{\bibfnamefont{A.}~\bibnamefont{Molinari}},
  \bibinfo{author}{\bibfnamefont{M.}~\bibnamefont{Monteno}},
  \bibinfo{author}{\bibfnamefont{M.}~\bibnamefont{Nardi}}, \bibnamefont{and}
  \bibinfo{author}{\bibfnamefont{F.}~\bibnamefont{Prino}},
  \bibinfo{journal}{Eur. Phys. J.} \textbf{\bibinfo{volume}{C71}},
  \bibinfo{pages}{1666} (\bibinfo{year}{2011})

\bibitem[{\citenamefont{Uphoff et~al.}(2012)\citenamefont{Uphoff, Fochler, Xu,
  and Greiner}}]{Uphoff:2012gb}
\bibinfo{author}{\bibfnamefont{J.}~\bibnamefont{Uphoff}},
  \bibinfo{author}{\bibfnamefont{O.}~\bibnamefont{Fochler}},
  \bibinfo{author}{\bibfnamefont{Z.}~\bibnamefont{Xu}}, \bibnamefont{and}
  \bibinfo{author}{\bibfnamefont{C.}~\bibnamefont{Greiner}},
  \bibinfo{journal}{Phys. Lett.} \textbf{\bibinfo{volume}{B717}},
  \bibinfo{pages}{430} (\bibinfo{year}{2012})

\bibitem[{\citenamefont{Lang et~al.}(2016)\citenamefont{Lang, van Hees,
  Steinheimer, Inghirami, and Bleicher}}]{Lang:2012cx}
\bibinfo{author}{\bibfnamefont{T.}~\bibnamefont{Lang}},
  \bibinfo{author}{\bibfnamefont{H.}~\bibnamefont{van Hees}},
  \bibinfo{author}{\bibfnamefont{J.}~\bibnamefont{Steinheimer}},
  \bibinfo{author}{\bibfnamefont{G.}~\bibnamefont{Inghirami}},
  \bibnamefont{and} \bibinfo{author}{\bibfnamefont{M.}~\bibnamefont{Bleicher}},
  \bibinfo{journal}{Phys. Rev.} \textbf{\bibinfo{volume}{C93}},
  \bibinfo{pages}{014901} (\bibinfo{year}{2016})

\bibitem[{\citenamefont{Song et~al.}(2015)\citenamefont{Song, Berrehrah,
  Cabrera, Torres-Rincon, Tolos, Cassing, and Bratkovskaya}}]{Song:2015sfa}
\bibinfo{author}{\bibfnamefont{T.}~\bibnamefont{Song}},
  \bibinfo{author}{\bibfnamefont{H.}~\bibnamefont{Berrehrah}},
  \bibinfo{author}{\bibfnamefont{D.}~\bibnamefont{Cabrera}},
  \bibinfo{author}{\bibfnamefont{J.~M.} \bibnamefont{Torres-Rincon}},
  \bibinfo{author}{\bibfnamefont{L.}~\bibnamefont{Tolos}},
  \bibinfo{author}{\bibfnamefont{W.}~\bibnamefont{Cassing}}, \bibnamefont{and}
  \bibinfo{author}{\bibfnamefont{E.}~\bibnamefont{Bratkovskaya}},
  \bibinfo{journal}{Phys. Rev.} \textbf{\bibinfo{volume}{C92}},
  \bibinfo{pages}{014910} (\bibinfo{year}{2015})

\bibitem[{\citenamefont{Song et~al.}(2016)\citenamefont{Song, Berrehrah,
  Cabrera, Cassing, and Bratkovskaya}}]{Song:2015ykw}
\bibinfo{author}{\bibfnamefont{T.}~\bibnamefont{Song}},
  \bibinfo{author}{\bibfnamefont{H.}~\bibnamefont{Berrehrah}},
  \bibinfo{author}{\bibfnamefont{D.}~\bibnamefont{Cabrera}},
  \bibinfo{author}{\bibfnamefont{W.}~\bibnamefont{Cassing}}, \bibnamefont{and}
  \bibinfo{author}{\bibfnamefont{E.}~\bibnamefont{Bratkovskaya}},
  \bibinfo{journal}{Phys. Rev.} \textbf{\bibinfo{volume}{C93}},
  \bibinfo{pages}{034906} (\bibinfo{year}{2016})

\bibitem[{\citenamefont{Das et~al.}(2014)\citenamefont{Das, Scardina, Plumari,
  and Greco}}]{Das:2013kea}
\bibinfo{author}{\bibfnamefont{S.~K.} \bibnamefont{Das}},
  \bibinfo{author}{\bibfnamefont{F.}~\bibnamefont{Scardina}},
  \bibinfo{author}{\bibfnamefont{S.}~\bibnamefont{Plumari}}, \bibnamefont{and}
  \bibinfo{author}{\bibfnamefont{V.}~\bibnamefont{Greco}},
  \bibinfo{journal}{Phys. Rev.} \textbf{\bibinfo{volume}{C90}},
  \bibinfo{pages}{044901} (\bibinfo{year}{2014})

\bibitem[{\citenamefont{Cao et~al.}(2015)\citenamefont{Cao, Qin, and
  Bass}}]{Cao:2015hia}
\bibinfo{author}{\bibfnamefont{S.}~\bibnamefont{Cao}},
  \bibinfo{author}{\bibfnamefont{G.-Y.} \bibnamefont{Qin}}, \bibnamefont{and}
  \bibinfo{author}{\bibfnamefont{S.~A.} \bibnamefont{Bass}},
  \bibinfo{journal}{Phys. Rev.} \textbf{\bibinfo{volume}{C92}},
  \bibinfo{pages}{024907} (\bibinfo{year}{2015})

\bibitem[{\citenamefont{Das et~al.}(2015)\citenamefont{Das, Scardina, Plumari,
  and Greco}}]{Das:2015ana}
\bibinfo{author}{\bibfnamefont{S.~K.} \bibnamefont{Das}},
  \bibinfo{author}{\bibfnamefont{F.}~\bibnamefont{Scardina}},
  \bibinfo{author}{\bibfnamefont{S.}~\bibnamefont{Plumari}}, \bibnamefont{and}
  \bibinfo{author}{\bibfnamefont{V.}~\bibnamefont{Greco}},
  \bibinfo{journal}{Phys. Lett.} \textbf{\bibinfo{volume}{B747}},
  \bibinfo{pages}{260} (\bibinfo{year}{2015})

\bibitem[{\citenamefont{Cao et~al.}(2018)\citenamefont{Cao, Luo, Qin, and
  Wang}}]{Cao:2017hhk}
\bibinfo{author}{\bibfnamefont{S.}~\bibnamefont{Cao}},
  \bibinfo{author}{\bibfnamefont{T.}~\bibnamefont{Luo}},
  \bibinfo{author}{\bibfnamefont{G.-Y.} \bibnamefont{Qin}}, \bibnamefont{and}
  \bibinfo{author}{\bibfnamefont{X.-N.} \bibnamefont{Wang}},
  \bibinfo{journal}{Phys. Lett.} \textbf{\bibinfo{volume}{B777}},
  \bibinfo{pages}{255} (\bibinfo{year}{2018})

\bibitem[{\citenamefont{Das et~al.}(2017{\natexlab{a}})\citenamefont{Das,
  Ruggieri, Scardina, Plumari, and Greco}}]{Das:2017dsh}
\bibinfo{author}{\bibfnamefont{S.~K.} \bibnamefont{Das}},
  \bibinfo{author}{\bibfnamefont{M.}~\bibnamefont{Ruggieri}},
  \bibinfo{author}{\bibfnamefont{F.}~\bibnamefont{Scardina}},
  \bibinfo{author}{\bibfnamefont{S.}~\bibnamefont{Plumari}}, \bibnamefont{and}
  \bibinfo{author}{\bibfnamefont{V.}~\bibnamefont{Greco}}, \bibinfo{journal}{J.
  Phys.} \textbf{\bibinfo{volume}{G44}}, \bibinfo{pages}{095102}
  (\bibinfo{year}{2017}{\natexlab{a}})

\bibitem[{\citenamefont{Sun et~al.}(2019)\citenamefont{Sun, Coci, Das, Plumari,
  Ruggieri, and Greco}}]{Sun:2019fud}
\bibinfo{author}{\bibfnamefont{Y.}~\bibnamefont{Sun}},
  \bibinfo{author}{\bibfnamefont{G.}~\bibnamefont{Coci}},
  \bibinfo{author}{\bibfnamefont{S.~K.} \bibnamefont{Das}},
  \bibinfo{author}{\bibfnamefont{S.}~\bibnamefont{Plumari}},
  \bibinfo{author}{\bibfnamefont{M.}~\bibnamefont{Ruggieri}}, \bibnamefont{and}
  \bibinfo{author}{\bibfnamefont{V.}~\bibnamefont{Greco}},
  \bibinfo{journal}{Phys. Lett.} \textbf{\bibinfo{volume}{B798}},
  \bibinfo{pages}{134933} (\bibinfo{year}{2019})

\bibitem[{\citenamefont{Plumari et~al.}(2011)\citenamefont{Plumari, Alberico,
  Greco, and Ratti}}]{Plumari:2011mk}
\bibinfo{author}{\bibfnamefont{S.}~\bibnamefont{Plumari}},
  \bibinfo{author}{\bibfnamefont{W.~M.} \bibnamefont{Alberico}},
  \bibinfo{author}{\bibfnamefont{V.}~\bibnamefont{Greco}}, \bibnamefont{and}
  \bibinfo{author}{\bibfnamefont{C.}~\bibnamefont{Ratti}},
  \bibinfo{journal}{Phys.Rev.} \textbf{\bibinfo{volume}{D84}},
  \bibinfo{pages}{094004} (\bibinfo{year}{2011})

\bibitem[{\citenamefont{Berrehrah et~al.}(2014)\citenamefont{Berrehrah,
  Gossiaux, Aichelin, Cassing, and Bratkovskaya}}]{Berrehrah:2014kba}
\bibinfo{author}{\bibfnamefont{H.}~\bibnamefont{Berrehrah}},
  \bibinfo{author}{\bibfnamefont{P.-B.} \bibnamefont{Gossiaux}},
  \bibinfo{author}{\bibfnamefont{J.}~\bibnamefont{Aichelin}},
  \bibinfo{author}{\bibfnamefont{W.}~\bibnamefont{Cassing}}, \bibnamefont{and}
  \bibinfo{author}{\bibfnamefont{E.}~\bibnamefont{Bratkovskaya}},
  \bibinfo{journal}{Phys. Rev.} \textbf{\bibinfo{volume}{C90}},
  \bibinfo{pages}{064906} (\bibinfo{year}{2014})

\bibitem[{\citenamefont{He et~al.}(2013)\citenamefont{He, Fries, and
  Rapp}}]{He:2012df}
\bibinfo{author}{\bibfnamefont{M.}~\bibnamefont{He}},
  \bibinfo{author}{\bibfnamefont{R.~J.} \bibnamefont{Fries}}, \bibnamefont{and}
  \bibinfo{author}{\bibfnamefont{R.}~\bibnamefont{Rapp}},
  \bibinfo{journal}{Phys. Rev. Lett.} \textbf{\bibinfo{volume}{110}},
  \bibinfo{pages}{112301} (\bibinfo{year}{2013})

\bibitem[{\citenamefont{He et~al.}(2012)\citenamefont{He, Fries, and
  Rapp}}]{He:2011qa}
\bibinfo{author}{\bibfnamefont{M.}~\bibnamefont{He}},
  \bibinfo{author}{\bibfnamefont{R.~J.} \bibnamefont{Fries}}, \bibnamefont{and}
  \bibinfo{author}{\bibfnamefont{R.}~\bibnamefont{Rapp}},
  \bibinfo{journal}{Phys. Rev.} \textbf{\bibinfo{volume}{C86}},
  \bibinfo{pages}{014903} (\bibinfo{year}{2012})

\bibitem[{\citenamefont{Cao et~al.}(2016)\citenamefont{Cao, Luo, Qin, and
  Wang}}]{Cao:2016gvr}
\bibinfo{author}{\bibfnamefont{S.}~\bibnamefont{Cao}},
  \bibinfo{author}{\bibfnamefont{T.}~\bibnamefont{Luo}},
  \bibinfo{author}{\bibfnamefont{G.-Y.} \bibnamefont{Qin}}, \bibnamefont{and}
  \bibinfo{author}{\bibfnamefont{X.-N.} \bibnamefont{Wang}},
  \bibinfo{journal}{Phys. Rev.} \textbf{\bibinfo{volume}{C94}},
  \bibinfo{pages}{014909} (\bibinfo{year}{2016})

\bibitem[{\citenamefont{Scardina et~al.}(2010)\citenamefont{Scardina, Di~Toro,
  and Greco}}]{SCARDINA}
\bibinfo{author}{\bibfnamefont{F.}~\bibnamefont{Scardina}},
  \bibinfo{author}{\bibfnamefont{M.}~\bibnamefont{Di~Toro}}, \bibnamefont{and}
  \bibinfo{author}{\bibfnamefont{V.}~\bibnamefont{Greco}},
  \bibinfo{journal}{Phys.Rev.} \textbf{\bibinfo{volume}{C82}},
  \bibinfo{pages}{054901} (\bibinfo{year}{2010}).

\bibitem[{\citenamefont{Liao and Shuryak}(2009)}]{Liao:2008dk}
\bibinfo{author}{\bibfnamefont{J.}~\bibnamefont{Liao}} \bibnamefont{and}
  \bibinfo{author}{\bibfnamefont{E.}~\bibnamefont{Shuryak}},
  \bibinfo{journal}{Phys. Rev. Lett.} \textbf{\bibinfo{volume}{102}},
  \bibinfo{pages}{202302} (\bibinfo{year}{2009})

\bibitem[{\citenamefont{Xu et~al.}(2015)\citenamefont{Xu, Liao, and
  Gyulassy}}]{Xu:2014tda}
\bibinfo{author}{\bibfnamefont{J.}~\bibnamefont{Xu}},
  \bibinfo{author}{\bibfnamefont{J.}~\bibnamefont{Liao}}, \bibnamefont{and}
  \bibinfo{author}{\bibfnamefont{M.}~\bibnamefont{Gyulassy}},
  \bibinfo{journal}{Chin. Phys. Lett.} \textbf{\bibinfo{volume}{32}},
  \bibinfo{pages}{092501} (\bibinfo{year}{2015})

\bibitem[{\citenamefont{Xu et~al.}(2018)\citenamefont{Xu, Bernhard, Bass,
  Nahrgang, and Cao}}]{Xu:2017obm}
\bibinfo{author}{\bibfnamefont{Y.}~\bibnamefont{Xu}},
  \bibinfo{author}{\bibfnamefont{J.~E.} \bibnamefont{Bernhard}},
  \bibinfo{author}{\bibfnamefont{S.~A.} \bibnamefont{Bass}},
  \bibinfo{author}{\bibfnamefont{M.}~\bibnamefont{Nahrgang}}, \bibnamefont{and}
  \bibinfo{author}{\bibfnamefont{S.}~\bibnamefont{Cao}},
  \bibinfo{journal}{Phys. Rev.} \textbf{\bibinfo{volume}{C97}},
  \bibinfo{pages}{014907} (\bibinfo{year}{2018})

\bibitem[{\citenamefont{Scardina et~al.}(2017)\citenamefont{Scardina, Das,
  Minissale, Plumari, and Greco}}]{Scardina:2017ipo}
\bibinfo{author}{\bibfnamefont{F.}~\bibnamefont{Scardina}},
  \bibinfo{author}{\bibfnamefont{S.~K.} \bibnamefont{Das}},
  \bibinfo{author}{\bibfnamefont{V.}~\bibnamefont{Minissale}},
  \bibinfo{author}{\bibfnamefont{S.}~\bibnamefont{Plumari}}, \bibnamefont{and}
  \bibinfo{author}{\bibfnamefont{V.}~\bibnamefont{Greco}},
  \bibinfo{journal}{Phys. Rev.} \textbf{\bibinfo{volume}{C96}},
  \bibinfo{pages}{044905} (\bibinfo{year}{2017})

\bibitem[{\citenamefont{Nahrgang et~al.}(2015)\citenamefont{Nahrgang, Aichelin,
  Bass, Gossiaux, and Werner}}]{Nahrgang:2014vza}
\bibinfo{author}{\bibfnamefont{M.}~\bibnamefont{Nahrgang}},
  \bibinfo{author}{\bibfnamefont{J.}~\bibnamefont{Aichelin}},
  \bibinfo{author}{\bibfnamefont{S.}~\bibnamefont{Bass}},
  \bibinfo{author}{\bibfnamefont{P.~B.} \bibnamefont{Gossiaux}},
  \bibnamefont{and} \bibinfo{author}{\bibfnamefont{K.}~\bibnamefont{Werner}},
  \bibinfo{journal}{Phys. Rev.} \textbf{\bibinfo{volume}{C91}},
  \bibinfo{pages}{014904} (\bibinfo{year}{2015})

\bibitem[{\citenamefont{Prado et~al.}(2017)\citenamefont{Prado,
  Noronha-Hostler, Katz, Suaide, Noronha, Munhoz, and
  Cosentino}}]{Prado:2016szr}
\bibinfo{author}{\bibfnamefont{C.~A.~G.} \bibnamefont{Prado}},
  \bibinfo{author}{\bibfnamefont{J.}~\bibnamefont{Noronha-Hostler}},
  \bibinfo{author}{\bibfnamefont{R.}~\bibnamefont{Katz}},
  \bibinfo{author}{\bibfnamefont{A.~A.~P.} \bibnamefont{Suaide}},
  \bibinfo{author}{\bibfnamefont{J.}~\bibnamefont{Noronha}},
  \bibinfo{author}{\bibfnamefont{M.~G.} \bibnamefont{Munhoz}},
  \bibnamefont{and} \bibinfo{author}{\bibfnamefont{M.~R.}
  \bibnamefont{Cosentino}}, \bibinfo{journal}{Phys. Rev.}
  \textbf{\bibinfo{volume}{C96}}, \bibinfo{pages}{064903}
  (\bibinfo{year}{2017})

\bibitem[{\citenamefont{Nahrgang et~al.}(2016)\citenamefont{Nahrgang, Aichelin,
  Gossiaux, and Werner}}]{Nahrgang:2016wig}
\bibinfo{author}{\bibfnamefont{M.}~\bibnamefont{Nahrgang}},
  \bibinfo{author}{\bibfnamefont{J.}~\bibnamefont{Aichelin}},
  \bibinfo{author}{\bibfnamefont{P.~B.} \bibnamefont{Gossiaux}},
  \bibnamefont{and} \bibinfo{author}{\bibfnamefont{K.}~\bibnamefont{Werner}},
  \bibinfo{journal}{J. Phys. Conf. Ser.} \textbf{\bibinfo{volume}{668}},
  \bibinfo{pages}{012024} (\bibinfo{year}{2016}).

\bibitem[{\citenamefont{Beraudo et~al.}(2019)\citenamefont{Beraudo, De~Pace,
  Monteno, Nardi, and Prino}}]{Beraudo:2018tpr}
\bibinfo{author}{\bibfnamefont{A.}~\bibnamefont{Beraudo}},
  \bibinfo{author}{\bibfnamefont{A.}~\bibnamefont{De~Pace}},
  \bibinfo{author}{\bibfnamefont{M.}~\bibnamefont{Monteno}},
  \bibinfo{author}{\bibfnamefont{M.}~\bibnamefont{Nardi}}, \bibnamefont{and}
  \bibinfo{author}{\bibfnamefont{F.}~\bibnamefont{Prino}},
  \bibinfo{journal}{Eur. Phys. J.} \textbf{\bibinfo{volume}{C79}},
  \bibinfo{pages}{494} (\bibinfo{year}{2019})

\bibitem[{\citenamefont{Katz et~al.}(2019)\citenamefont{Katz, Prado,
  Noronha-Hostler, Noronha, and Suaide}}]{Katz:2019fkc}
\bibinfo{author}{\bibfnamefont{R.}~\bibnamefont{Katz}},
  \bibinfo{author}{\bibfnamefont{C.~A.~G.} \bibnamefont{Prado}},
  \bibinfo{author}{\bibfnamefont{J.}~\bibnamefont{Noronha-Hostler}},
  \bibinfo{author}{\bibfnamefont{J.}~\bibnamefont{Noronha}}, \bibnamefont{and}
  \bibinfo{author}{\bibfnamefont{A.~A.~P.} \bibnamefont{Suaide}}
  (\bibinfo{year}{2019})

\bibitem[{\citenamefont{Plumari et~al.}(2015)\citenamefont{Plumari, Guardo,
  Scardina, and Greco}}]{Plumari:2015cfa}
\bibinfo{author}{\bibfnamefont{S.}~\bibnamefont{Plumari}},
  \bibinfo{author}{\bibfnamefont{G.~L.} \bibnamefont{Guardo}},
  \bibinfo{author}{\bibfnamefont{F.}~\bibnamefont{Scardina}}, \bibnamefont{and}
  \bibinfo{author}{\bibfnamefont{V.}~\bibnamefont{Greco}},
  \bibinfo{journal}{Phys. Rev.} \textbf{\bibinfo{volume}{C92}},
  \bibinfo{pages}{054902} (\bibinfo{year}{2015})

\bibitem[{\citenamefont{Plumari}(2019)}]{Plumari:2019gwq}
\bibinfo{author}{\bibfnamefont{S.}~\bibnamefont{Plumari}},
  \bibinfo{journal}{Eur. Phys. J.} \textbf{\bibinfo{volume}{C79}},
  \bibinfo{pages}{2} (\bibinfo{year}{2019}).

\bibitem[{\citenamefont{Das et~al.}(2017{\natexlab{b}})\citenamefont{Das,
  Plumari, Chatterjee, Alam, Scardina, and Greco}}]{Das:2016cwd}
\bibinfo{author}{\bibfnamefont{S.~K.} \bibnamefont{Das}},
  \bibinfo{author}{\bibfnamefont{S.}~\bibnamefont{Plumari}},
  \bibinfo{author}{\bibfnamefont{S.}~\bibnamefont{Chatterjee}},
  \bibinfo{author}{\bibfnamefont{J.}~\bibnamefont{Alam}},
  \bibinfo{author}{\bibfnamefont{F.}~\bibnamefont{Scardina}}, \bibnamefont{and}
  \bibinfo{author}{\bibfnamefont{V.}~\bibnamefont{Greco}},
  \bibinfo{journal}{Phys. Lett.} \textbf{\bibinfo{volume}{B768}},
  \bibinfo{pages}{260} (\bibinfo{year}{2017}{\natexlab{b}})

\bibitem[{\citenamefont{Plumari et~al.}(2018)\citenamefont{Plumari, Minissale,
  Das, Coci, and Greco}}]{Plumari:2017ntm}
\bibinfo{author}{\bibfnamefont{S.}~\bibnamefont{Plumari}},
  \bibinfo{author}{\bibfnamefont{V.}~\bibnamefont{Minissale}},
  \bibinfo{author}{\bibfnamefont{S.~K.} \bibnamefont{Das}},
  \bibinfo{author}{\bibfnamefont{G.}~\bibnamefont{Coci}}, \bibnamefont{and}
  \bibinfo{author}{\bibfnamefont{V.}~\bibnamefont{Greco}},
  \bibinfo{journal}{Eur. Phys. J.} \textbf{\bibinfo{volume}{C78}},
  \bibinfo{pages}{348} (\bibinfo{year}{2018})

\bibitem[{\citenamefont{Ferini et~al.}(2009)\citenamefont{Ferini, Colonna,
  Di~Toro, and Greco}}]{Ferini:2008he}
\bibinfo{author}{\bibfnamefont{G.}~\bibnamefont{Ferini}},
  \bibinfo{author}{\bibfnamefont{M.}~\bibnamefont{Colonna}},
  \bibinfo{author}{\bibfnamefont{M.}~\bibnamefont{Di~Toro}}, \bibnamefont{and}
  \bibinfo{author}{\bibfnamefont{V.}~\bibnamefont{Greco}},
  \bibinfo{journal}{Phys.Lett.} \textbf{\bibinfo{volume}{B670}},
  \bibinfo{pages}{325} (\bibinfo{year}{2009})

\bibitem[{\citenamefont{Ruggieri et~al.}(2014)\citenamefont{Ruggieri, Scardina,
  Plumari, and Greco}}]{Ruggieri:2013ova}
\bibinfo{author}{\bibfnamefont{M.}~\bibnamefont{Ruggieri}},
  \bibinfo{author}{\bibfnamefont{F.}~\bibnamefont{Scardina}},
  \bibinfo{author}{\bibfnamefont{S.}~\bibnamefont{Plumari}}, \bibnamefont{and}
  \bibinfo{author}{\bibfnamefont{V.}~\bibnamefont{Greco}},
  \bibinfo{journal}{Phys.Rev.} \textbf{\bibinfo{volume}{C89}},
  \bibinfo{pages}{054914} (\bibinfo{year}{2014})

\bibitem[{\citenamefont{Qin et~al.}(2010)\citenamefont{Qin, Petersen, Bass, and
  Muller}}]{Qin:2010pf}
\bibinfo{author}{\bibfnamefont{G.-Y.} \bibnamefont{Qin}},
  \bibinfo{author}{\bibfnamefont{H.}~\bibnamefont{Petersen}},
  \bibinfo{author}{\bibfnamefont{S.~A.} \bibnamefont{Bass}}, \bibnamefont{and}
  \bibinfo{author}{\bibfnamefont{B.}~\bibnamefont{Muller}},
  \bibinfo{journal}{Phys. Rev.} \textbf{\bibinfo{volume}{C82}},
  \bibinfo{pages}{064903} (\bibinfo{year}{2010})

\bibitem[{\citenamefont{Petersen et~al.}(2010)\citenamefont{Petersen, Qin,
  Bass, and Muller}}]{Petersen:2010cw}
\bibinfo{author}{\bibfnamefont{H.}~\bibnamefont{Petersen}},
  \bibinfo{author}{\bibfnamefont{G.-Y.} \bibnamefont{Qin}},
  \bibinfo{author}{\bibfnamefont{S.~A.} \bibnamefont{Bass}}, \bibnamefont{and}
  \bibinfo{author}{\bibfnamefont{B.}~\bibnamefont{Muller}},
  \bibinfo{journal}{Phys.Rev.} \textbf{\bibinfo{volume}{C82}},
  \bibinfo{pages}{041901} (\bibinfo{year}{2010})

\bibitem[{\citenamefont{Cacciari et~al.}(2012)\citenamefont{Cacciari, Frixione,
  Houdeau, Mangano, Nason, and Ridolfi}}]{Cacciari:2012ny}
\bibinfo{author}{\bibfnamefont{M.}~\bibnamefont{Cacciari}},
  \bibinfo{author}{\bibfnamefont{S.}~\bibnamefont{Frixione}},
  \bibinfo{author}{\bibfnamefont{N.}~\bibnamefont{Houdeau}},
  \bibinfo{author}{\bibfnamefont{M.~L.} \bibnamefont{Mangano}},
  \bibinfo{author}{\bibfnamefont{P.}~\bibnamefont{Nason}}, \bibnamefont{and}
  \bibinfo{author}{\bibfnamefont{G.}~\bibnamefont{Ridolfi}},
  \bibinfo{journal}{JHEP} \textbf{\bibinfo{volume}{10}}, \bibinfo{pages}{137}
  (\bibinfo{year}{2012})

\bibitem[{\citenamefont{Combridge}(1979)}]{Combridge:1978kx}
\bibinfo{author}{\bibfnamefont{B.}~\bibnamefont{Combridge}},
  \bibinfo{journal}{Nucl.Phys.} \textbf{\bibinfo{volume}{B151}},
  \bibinfo{pages}{429} (\bibinfo{year}{1979}).

\bibitem[{\citenamefont{Peterson et~al.}(1983)\citenamefont{Peterson,
  Schlatter, Schmitt, and Zerwas}}]{Peterson:1982ak}
\bibinfo{author}{\bibfnamefont{C.}~\bibnamefont{Peterson}},
  \bibinfo{author}{\bibfnamefont{D.}~\bibnamefont{Schlatter}},
  \bibinfo{author}{\bibfnamefont{I.}~\bibnamefont{Schmitt}}, \bibnamefont{and}
  \bibinfo{author}{\bibfnamefont{P.~M.} \bibnamefont{Zerwas}},
  \bibinfo{journal}{Phys. Rev.} \textbf{\bibinfo{volume}{D27}},
  \bibinfo{pages}{105} (\bibinfo{year}{1983}).

\bibitem[{\citenamefont{Sirunyan et~al.}(2018)}]{Sirunyan:2017plt}
\bibinfo{author}{\bibfnamefont{A.~M.} \bibnamefont{Sirunyan}}
  \bibnamefont{et~al.} (\bibinfo{collaboration}{CMS}), \bibinfo{journal}{Phys.
  Rev. Lett.} \textbf{\bibinfo{volume}{120}}, \bibinfo{pages}{202301}
  (\bibinfo{year}{2018})

\bibitem[{\citenamefont{Adam et~al.}(2019)}]{Adam:2019hpq}
\bibinfo{author}{\bibfnamefont{J.}~\bibnamefont{Adam}} \bibnamefont{et~al.}
  (\bibinfo{collaboration}{STAR}) (\bibinfo{year}{2019}), \eprint{1910.14628}.

\bibitem[{\citenamefont{Vermunt}(2019)}]{Vermunt:2019ecg}
\bibinfo{author}{\bibfnamefont{L.}~\bibnamefont{Vermunt}}
  (\bibinfo{collaboration}{ALICE}), in \emph{\bibinfo{booktitle}{{2019 European
  Physical Society Conference on High Energy Physics (EPS-HEP2019) Ghent,
  Belgium, July 10-17, 2019}}} (\bibinfo{year}{2019})

\bibitem[{\citenamefont{Gardim et~al.}(2012)\citenamefont{Gardim, Grassi,
  Luzum, and Ollitrault}}]{Gardim:2011xv}
\bibinfo{author}{\bibfnamefont{F.~G.} \bibnamefont{Gardim}},
  \bibinfo{author}{\bibfnamefont{F.}~\bibnamefont{Grassi}},
  \bibinfo{author}{\bibfnamefont{M.}~\bibnamefont{Luzum}}, \bibnamefont{and}
  \bibinfo{author}{\bibfnamefont{J.-Y.} \bibnamefont{Ollitrault}},
  \bibinfo{journal}{Phys.Rev.} \textbf{\bibinfo{volume}{C85}},
  \bibinfo{pages}{024908} (\bibinfo{year}{2012})

\bibitem[{\citenamefont{Chaudhuri et~al.}(2013)\citenamefont{Chaudhuri, Haque,
  Roy, and Mohanty}}]{Chaudhuri:2012mr}
\bibinfo{author}{\bibfnamefont{A.~K.} \bibnamefont{Chaudhuri}},
  \bibinfo{author}{\bibfnamefont{M.~R.} \bibnamefont{Haque}},
  \bibinfo{author}{\bibfnamefont{V.}~\bibnamefont{Roy}}, \bibnamefont{and}
  \bibinfo{author}{\bibfnamefont{B.}~\bibnamefont{Mohanty}},
  \bibinfo{journal}{Phys. Rev.} \textbf{\bibinfo{volume}{C87}},
  \bibinfo{pages}{034907} (\bibinfo{year}{2013})

\bibitem[{\citenamefont{Niemi et~al.}(2013)\citenamefont{Niemi, Denicol,
  Holopainen, and Huovinen}}]{Niemi:2012aj}
\bibinfo{author}{\bibfnamefont{H.}~\bibnamefont{Niemi}},
  \bibinfo{author}{\bibfnamefont{G.}~\bibnamefont{Denicol}},
  \bibinfo{author}{\bibfnamefont{H.}~\bibnamefont{Holopainen}},
  \bibnamefont{and} \bibinfo{author}{\bibfnamefont{P.}~\bibnamefont{Huovinen}},
  \bibinfo{journal}{Phys.Rev.} \textbf{\bibinfo{volume}{C87}},
  \bibinfo{pages}{054901} (\bibinfo{year}{2013})

\bibitem[{\citenamefont{Aad et~al.}(2013)}]{Aad:2013xma}
\bibinfo{author}{\bibfnamefont{G.}~\bibnamefont{Aad}} \bibnamefont{et~al.}
  (\bibinfo{collaboration}{ATLAS Collaboration}), \bibinfo{journal}{JHEP}
\textbf{\bibinfo{volume}{1311}}, \bibinfo{pages}{183} (\bibinfo{year}{2013})

\end{thebibliography}
%

\end{document}